\documentstyle[11pt,paspconf,epsf]{article}
\def\ref{\par\noindent\hangindent=1truecm}
\font\piedi=cmr8

\catcode`\@=11
\def\gsim{\ifmmode{\mathrel{\mathpalette\@versim>}}
    \else{$\mathrel{\mathpalette\@versim>}$}\fi}
\def\lsim{\ifmmode{\mathrel{\mathpalette\@versim<}}
    \else{$\mathrel{\mathpalette\@versim<}$}\fi}
\def\@versim#1#2{\lower 2.9truept \vbox{\baselineskip 0pt \lineskip 
    0.5truept \ialign{$\m@th#1\hfil##\hfil$\crcr#2\crcr\sim\crcr}}}
\catcode`\@=12
\def\pn{\par\noindent}
\def\lsun{\hbox{$L_\odot$}}
\def\lb{\hbox{$L_{\rm B}$}}
\def\msun{\hbox{$M_\odot$}}
\def\IMLR{Fe$M/L$}
\def\micm{\hbox{$M_{\rm ICM}$}}
\def\mfecm{\hbox{$M_{\rm Fe}^{\rm ICM}$}}

\def\mfes{\hbox{$M_{\rm Fe}^*$}}

\def\zfes{\hbox{$Z^{\rm Fe}_*$}}
\def\zfecm{\hbox{$Z^{\rm Fe}_{\rm ICM}$}}
\def\ho{\hbox{$H_\circ$}}
\def\h50{\hbox{$\ho /50$}}
\begin{document}

\title{The Main Epoch of Metal Production in the Universe}
\author{Alvio Renzini}

\affil{European Southern Observatory, Garching b. M\"unchen, Germany}




\begin{abstract}
Clusters of galaxies allow a direct estimate of the metallicity and
metal production yield on the largest scale so far. It is argued that
cluster metallicity ($\sim 1/3$ solar) should be taken as
representative of the low-$z$ universe as a whole.  There is now
compelling evidence that the bulk of stars not only in cluster
ellipticals but also in field ellipticals and bulges formed at high 
redshifts ($z\gsim 3$). Since such stars account for at least $\sim
30\%$ of the baryons now locked into stars, it is argued that at least
$30\%$ of stars and metals formed before $z\simeq 3$, and
correspondingly the
metallicity of the universe at $z=3$ is predicted to be $\sim 1/10$ solar.

\keywords{elliptical galaxies, clusters of galaxies, chemical
evolution, high redshift galaxies}
            
\end{abstract}      


\section{Introduction}

My aim with this paper is to use the local, {\it fossil evidence}
(i.e. the global metallicity and stellar ages at $z\simeq 0$) to get
clues and set constraints on the past metal production and star
formation rate (SFR) in the
universe. This attempt (see also Rich 1997) is therefore 
complementary to current efforts
to add more and more accurate data points to the global  SFR vs
redshift diagram (the Madau diagram, for short; Madau et al. 1996,
1997; Madau, these proceedings).

The paper is organized as follows. Section 2 presents the current
evidence for the chemical composition of local clusters of galaxies,
at low redshift. Section 3 discusses to which extent clusters are 
representative of the
low redshift universe as a whole. In Section 4 a plea is presented for
the bulk of stars in galactic spheroids (i.e. ellipticals and bulges
alike) being very old, formed at high redshift, hence for a fair
fraction of the metals we see at $z\simeq 0$ having been produced at
high $z$. I will conclude that this scenario favors high SFRs at
high-$z$, with the SFR peaking at $z\gsim 2$. Some of these topics are
expanded upon in Renzini (1997, hereafter R97).

\begin{figure}
\vskip-5truecm
\plotone{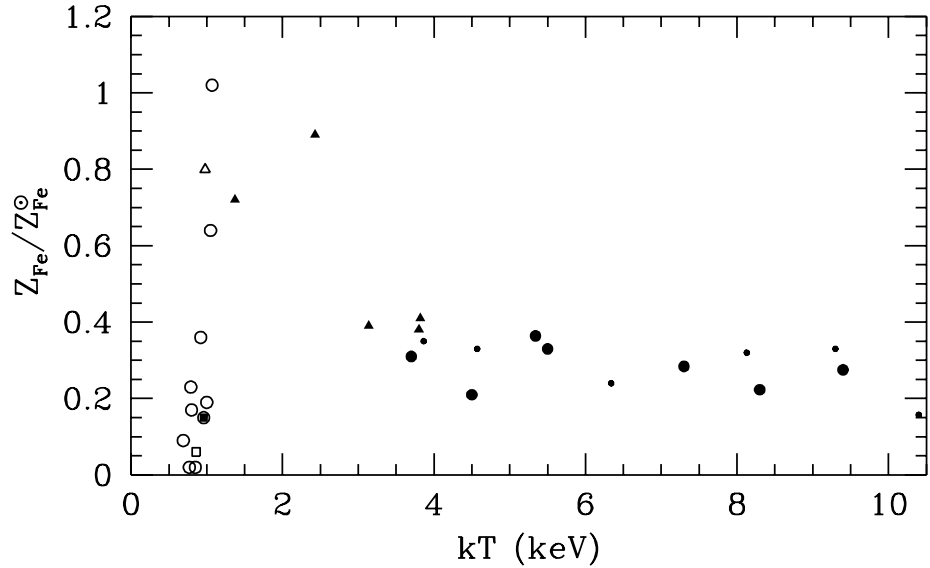}
\vskip-1truecm
\caption{\piedi The iron abundance in the ICM as a function of ICM temperature
for a sample of clusters and groups, including of six clusters at
moderately high redshift with $<z>\simeq 0.33$, represented by small
filled circles (from R97).}
\end{figure}
 
\section{Clusters as Archives of the Past Star and Metal Production} 
Theoretical simulations predict that the baryon fraction of
clusters
cannot change appreciably in the course of their evolution (White et
al. 1993). This is to say that -- unlike individual
galaxies -- clusters are good examples of a {\it closed box}. Metals
are ejected by galaxies but retained by clusters. Moreover, as
the baryon fraction remains nearly constant no extra-dilution of
metals takes place, and eventually we find confined in the same place
all the dark matter, all the baryons, all the galaxies, and all the
metals that have participated in the play. Hence, clusters are good {\it
archives} of their past star formation and metal production history
(Cavaliere, private communication).

Metals in clusters are partly spread through their intracluster medium
(ICM), partly locked into galaxies and stars. By comparison, the mass
of metals 
ISM of galaxies is negligible. ICM abundances can be obtained from X-ray 
observations, while optical observations combined to population
synthesis models provide estimates for the metallicity of the stellar
component of galaxies.

\subsection{Iron and $\alpha$-Elements in the Intracluster Medium}

The best known ICM abundance is that of iron. It comes from the
so-called iron-K emission complex at $\sim 7$ keV, prominent in the
X-ray spectrum of clusters. Fig. 1 shows the
iron abundance of clusters and groups as a function of ICM
temperature. Still as a function of ICM temperature,
Fig. 2 shows instead the iron-mass-to-ligh-ratio(\IMLR) of the cluster ICM,
measured as the ratio $\mfecm/\lb$ of the total iron mass in
the ICM over the total $B$-band luminosity of the galaxies in the cluster.

\begin{figure}
\vskip-5truecm
\plotone{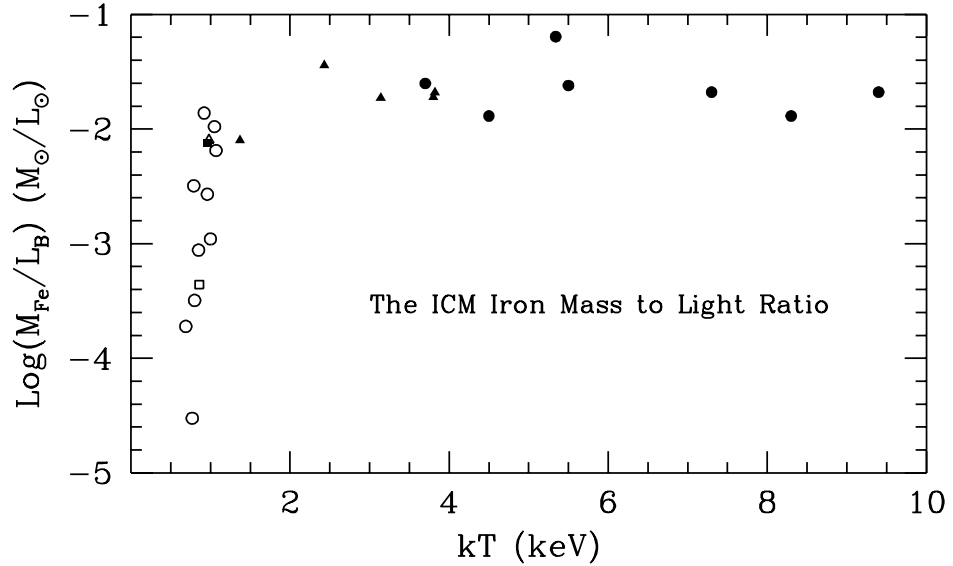}
\vskip-1truecm
\caption{\piedi The iron mass to light ratio  of  the  ICM of clusters
and groups (for $\ho=50$) as a function of the ICM temperature (from
R97).}
\end{figure}

The drop of the derived \IMLR \ in poor clusters
 and groups (i.e. for $kT\lsim 2$ keV) can be traced back
to a drop in both factors entering in its definition, i.e., in 
the iron abundance {\it and} in the ICM mass to light ratio. It is not
clear whether this is a real effect, signalling that groups are
not closed boxes, or that diagnostic problems are present due to iron
being derived from the iron-L instead than from the iron-K complex (cf. R97).
I will not further discuss of $kT\lsim 2$ keV objects.

What emerges from Fig. 1 and 2 is that both the iron abundance and the
\IMLR \ in rich clusters ($kT\gsim 2$ keV) are constant,
i.e. independent of cluster temperature, hence of cluster richness and
optical luminosity that are correlated quantities. In practice,
$\zfecm =0.3\pm 0.1$ solar, and $\mfecm/\lb = (0.02\pm 0.01)$.
The  simplest interpretation of all this is that clusters did not lose
iron (hence baryons), nor acquired pristine baryonic material, and
that the conversion of baryonic gas to stars and galaxies has
proceeded with the same efficiency and stellar IMF in all clusters (cf. R97).
The theoretical predictions on the baryon fraction in clusters
(cf. Section 1) find a nice support from these evidences. 

Besides iron, X-ray observations allow to measure the abundance of
other elements in the ICM, especially of the $\alpha$-elements such as
O, Ne, Mg, and Si, with {\it ASCA} having superceded any
previous attempt in this respect. A fairly high $\alpha$-element
enhancement, with
$<\![\alpha$/Fe]$>\simeq +0.4$, was initially reported (Mushotzky
1994), but more recently Mushotzky et al. (1996) have revised down to
$<\![\alpha$/Fe]$>\simeq +0.2$ this estimate (taking a global average
for O, Ne, Mg, Si, and Fe). This may still suggest a modest
$\alpha$-element enhancement, with the ICM enrichment being dominated
by SNII products.

However, Ishimaru \& Arimoto (1997) have recently pointed out that the
small apparent $\alpha$-element enhancement in the ICM comes from Mushotzky et
al. (1996)
having assumed reference solar abundances from ``photospheric'' model
atmosphere analysis. The result is different if one uses the ``meteoritic'' 
iron abundance instead, which is $\sim 0.16$
dex  lower than the photospheric value. Since the meteoritic value  is
now generally adopted for the solar iron abundance, one can conclude that
there is virtually no $\alpha$-element enhancement at all in
the ICM (formally $<\![\alpha$/Fe]$>\simeq +0.04\;\pm\sim 0.2$).
It is eventually quite reassuring to find that clusters of galaxies
are {\it solar} as far as the
elemental ratios are concerned, which argues for stellar 
nucleosynthesis having proceeded in quite the same way in the solar 
neighborhood as well as at the galaxy cluster scale. Specifically,
this implies a similar ratio of Type Ia to Type II SNs, as well as
a similar IMF (R97). If not else, this will help limiting the number of free
parameters to play with.

\subsection{The ICM-Galaxies Iron Share}

In clusters of galaxies part of iron resides in stars, part in the
ICM, and the global
iron
abundance of the whole cluster is then given by:
$$Z_{\rm CL}^{\rm Fe}={\zfecm\micm + \zfes M_* \over \micm + M_*}=
      {5.5\zfecm h^{-5/2} + \zfes h^{-1}\over 5.5h^{-5/2} +
      h^{-1}},\eqno(1)$$
where $\zfes$ is the average abundance of stars in galaxies and $ M_*$
is the mass in stars. For the second equality 
I have assumed as prototypical the Coma cluster values
adopted by White et al. (1993): $\micm\simeq 5.5\times
10^{13}h^{-5/2}\msun$ and $M_*\simeq 10^{13}h^{-1}\msun$.
With $\zfecm=0.3$ solar and $\zfes=1$ solar,
equation (1) gives a global cluster abundance of 0.34,
0.37, and 0.41 times solar, respectively for $h=0.5$, 0.75, and 1. 
Under the same assumptions, the ratio of the iron mass in the ICM to
      the iron mass locked into stars is:
$${\zfecm\micm\over\zfes M_*}\simeq 1.65 h^{-3/2},\eqno(2)$$ or 4.6,
2.5, and 1.65, respectively for $h=0.5$, 0.75, and 1. Note that with
the adopted values for the quantities in equation (2) most of the iron
is in the ICM, rather than now locked into stars, especially for low
values of $\ho$. These estimates could be somewhat decreased if
clusters contain a sizable population of stars not bound to censed
individual galaxies, if the average iron abundance in stars is
supersolar (luminosity-weighted determinations underestimate true
abundances, Greggio 1997), or if the galaxy $M_*/L$ ratio is higher
than adopted here, i.e., $<\!M_*/\lb\!>=6.4h$ (White et al. 1993).
However, the bottom line is that there are at least as much metals inside
cluster galaxies, as there are out of them in the ICM. This must be
taken as a strong constraint when modelling the chemical evolution of galaxies:
clearly they do not evolve as a closed box, and outflows must play a
leading role.

With the adopted masses and iron abundances for the two baryonic
components  one can also evaluate the total
cluster \IMLR:
$${\mfecm +\mfes\over\lb}\simeq 1.3\times 10^{-2}(1.65\, h^{-1/2}+h)
\; (\msun/\lsun),\eqno(3)$$
or \IMLR=0.037 or 0.034 $\msun/\lsun$, respectively for $h=0.5$ and
1. The total
\IMLR \ is therefore fairly insensitive to the adopted distance scale.
Simple calculations (cf. Renzini et al. 1993)
show that to reproduce this value one needs either
a fairly flat IMF ($x\simeq 0.9$) if all iron is attributed to SNII's,
or a major contribution from SNIa's, if one adopts a Salpeter IMF
($x=1.35$). The former option
dictates a substantial $\alpha$-element enhancement, similar to the
values observed in the Galactic halo ([$\alpha$/Fe]$\simeq +0.5$).
The latter option instead predicts near solar proportions for the
cluster as a whole. The evidence presented in Section 2.1  favors of
the second option. From the near solar proportions of cluster
abundances one obtains the total metal mass to light ratio of a typical
cluster as $M_{\rm Z}/\lb\simeq 10\times M_{\rm Fe}/\lb\simeq  0.3\pm 0.1\;
(\msun/\lsun)$.

It is worth noting that this is an interesting estimate of the metal
yield of stellar populations that is fully empirical. Following
Tinsley (1980) the metal yield is usually defined per unit mass of
stars, a quantity which theoretical analog depends on the poorly known
low mass end of the
IMF. The estimate above gives instead the yield per unit luminosity of
present day cluster galaxies, a quantity that depends on the IMF only
for $M\gsim\msun$.  Theoretical mass-related yields have been recently
estimated by Thomas et al. (1997) based on massive star models by Woosley \&
Weaver (1995) and Thielemann et al. (1996). These yields can be purged
from their mass dependence, and transformed into luminosity-related
yields. For this purpose I assume an age of 15 Gyr for the bulk of
stars in clusters (cf. Section 4), and use the luminosity-IMF
normalization from
(Renzini 1994): i.e. $\psi(M)=AM^{-(1+x)}$ for the IMF, one has 
$A\simeq 3.0\lb$. Thus, theoretical yields turn out to be $M_{\rm
Z}/\lb=0.08$, 0.24, and $0.33\; \msun/\lsun$, respectively for
$x=1.7$, 1.35, and 1.00, which compares to $M_{\rm Z}/\lb\simeq 0.3\pm
0.1 \msun/\lsun$ for the empirical cluster value. One can conclude
that current stellar yields do not require a very flat IMF to account
for the cluster metals.

\section{Clusters vs Field}

A critical issue is to what extent the cluster global metallicity, and
the ICM to galaxies iron share are representative of the low$-z$
universe as a whole. For example, Madau et al. (1996) adopt $\ho=50$, a stellar
mass density parameter $\Omega_*=0.0036$, a baryon mass density
parameter $\Omega_{\rm b}=0.05$, an average solar metallicity for the
stars, and a negligible metal content for the intergalactic medium (IGM),
that comprises the vast majority of the baryons. With these
assumptions the metallicity of the present day universe is $\sim
1\times 0.0036/0.05= 0.07$ solar, or $\sim 5$ times lower than the measured
value in clusters of galaxies. In the same frame,  the fraction of baryons in
galaxies (stars) is $\sim 7\%$, which compares to  $\sim
1/(1+5.5h^{-3/2})$ in clusters, or $\sim 6\%$ and $\sim 10\%$,
respectively for $h=0.5$ and 0.75.
Therefore, it appears that the efficiency of baryon conversion into
galaxies and stars adopted  by Madau et
al. (1996) is nearly the same as that observed in clusters, which
supports the notion of clusters being representative of the low-$z$
universe $(\Omega_*/\Omega_{\rm b})_\circ$. 
The metallicity of the clusters is however $\sim 5$ times higher
than the metallicity of the low-$z$ universe adopted by Madau et al.. 
The difference  comes from having assumed
 the IGM to be devoid metals, hence assuming field galaxies losing 
a negligible amount of metals, contrary to cluster galaxies which
instead
appear to have lost a major fraction of the metals they have
produced. Since the average stellar metallicity
is assumed to be solar in both clusters and field, this also implies a
factor $\sim 5$
lower  efficiency in metal production per unit mass turned into stars
(yield), compared to galaxy clusters.

Assuming to be real, such drastic differences between the behavior of
 galaxies and stellar populations in  clusters and in the field would require
 quite contrived explanations (R97). More attractive for its
 simplicity appears to be the
alternative according to which no major difference exists between
 field and
clusters, and  the global
metallicity of the present day universe is nearly the same as
that observed in galaxy clusters, i.e., $\sim 1/3$ times solar.
If so, there should be a comparable share of metals in the field IGM,
as there is in the cluster ICM, i.e., most of the metals should reside
in the IGM rather than within field galaxies (R97).

\section{The Main Epoch of  Metal Production and Star Formation}

In the previous section I have argued that the present metallicity
of the universe, hence the global, time-averaged
rate of metal production in are likely to be $\sim 5$
times higher than was adopted  by Madau et al. (1996). However, contrary
to Mushotzky \& Loewenstein (1997), this does not
necessarily imply that the global, time-averaged SFR  was also
underestimated by the same factor, because Madau et
al. have adopted for the baryon to stars conversion efficiency of the
general field precisely the same value found in clusters. 
Therefore, rather than to an higher average SFR one can appeal to
an higher metal yield, actually, just about the same yield empirically
determined for the clusters. 

Now, what is the main epoch of metal production in the
universe? or, equivalently, when the cosmic SFR has reached its peak
value?  The most straightforward approach to answer these questions is
certainly offered by the direct determination of the global SFR as a
function of redshift (the
Madau diagram). However, as well known, direct determinations of the global
SFR at high redshift encounters two major difficulties: 1) dust
obscuration, which on average may be higher than in low redshift
galaxies since star formation may predominantly take  place in major dust
enshrouded, {\it bulge-forming} starbursts, rather than more
quiescently as later in disks and
irregulars; and 2) alternatively, a sizable fraction of the global SFR
may take place in small entities below detection threshold, as 
predicted by current hierarchical models (e.g. Kaufmann 1996). Therefore, the
complementary approach based on the fossil record at low redshift
will provide a vision of the early universe from a different point of
view, hence subject to different biases and uncertainties. Ultimately, the two
approaches will have to lead to converging results. 

It is now well established by several independent lines of evidence
that the bulk of stars in cluster ellipticals formed at high redshift,
i.e., $z\gsim 3$ (an extensive set of references is not reported here
for lack of space, but can be found in R97). 
This important conclusion comes from the tightness of
the color$-\sigma$ relation of ellipticals in nearby clusters, the
tightness of the  distributions of cluster ellipticals
about their fundamental plane at low and high redshift,
the tightness of the Mg$_2-\sigma$ relation, again at low as well as high
redshift, the tightness of the color-magnitude relation of cluster
ellipticals at high redshift, and the small color and fundamental
plane evolution out to $z\simeq 0.5$. 

This {\it old age} conclusion can be 
generalized to include field ellipticals as well as the bulges of
spirals, and state  that the bulk of stars in galactic spheroids
formed at high redshift, i.e.  $z\gsim 2-3$. Indeed, field ellipticals
appear to follow the same Mg$_2-\sigma$ relation of their cluster
analogs (Bernardi et al. 1998), while galactic bulges follow the same
Mg$_2$-luminosity (and Mg$_2-\sigma$) relation of ellipticals
(Jablonka et al. 1996). Moreover, stellar photometry shows that
the bulk of stars in the Galactic bulge are as old as the 
Galactic halo, or $\sim 15$ Gyr (Ortolani et al. 1995), and that there
is no evidence for an intermediate age population in the bulge of M31
and its elliptical satellites M32 and NGC 147 (Renzini 1998).
The Milky Way and M31 reside in a small, spiral dominated group and
yet their bulges are $\sim 15$ Gyr old. It seems reasonable to
generalize this to virtually all bulges, given also the strong
similarity of bulges and ellipticals.
 
Acoording to Schechter \& Dressler (1987), bulges account for nearly
the same star mass as disks, while the star mass in E/S0 galaxies is
about twice that in spirals (Persic \& Salucci 1992). Hence, it seems 
legitimate  to conclude that galactic spheroids contain $\sim 30-50\%$ of
the baryons now locked into stars.  Therefore, if the bulk of stars in
spheroids formed at $z\gsim 3$, and they account for $\gsim 30\%$ of
the stellar mass at $z=0$, putting two and two together:
\smallskip\pn$\bullet$ 
Some 30\% of all stars have formed at $z\gsim 3$,
\smallskip\pn$\bullet$
hence, $\sim 30\%$ of all the metals have been produced at $z\gsim 3$,
\smallskip\pn$\bullet$
hence, the metallicity of the universe at $z=3$ should be $\sim 1/3$
of its present day value  ($\sim 1/3$ solar), i.e., $\sim 1/10$
solar, or $Z(z=3)\simeq 0.002$.

At first sight these inferences from the local, fossil evidence appear
to conflict with some model predictions, and with some of the current
 interpretations of the direct observations at $z\gsim
3$.  For example, the standard CDM model of Cole et al. (1994) and
Baugh et al. (1997) predicts that only $\lsim 5\%$ of stars have
formed by $z=3$. Tuning the free parameters of this specific CDM model does not
seem to work without violating other observational constraints (Frenk,
this conference). It seems that more drastic changes are required to
produce $\sim 30\%$ of the stars by $z=3$, perhaps appealing to 
isocurvature models designed to allow the early assembly of large galaxies 
(Peebles 1997).

As far as direct observations are concerned, the fossil evidence
supports the interpretation as {\it lower limits} of the SFR estimates
at $z\gsim 3$ (e.g. Madau et al. 1996).  Of the two models presented
by Madau et al. (1997) the one in which the SFR slowly increases all
the way to $z=5$ is favored. It remains to be ascertained 
whether the undetected star formation activity is obscured by dust, or
dispersed in small fragments below the  threshold of the Hubble
Deep Field.

Finally, the predicted global metallicity at $z=3$ ($\sim 1/10$ solar,
or more) is subject to observational test via the damped Ly$_\alpha$
systems (DLA). These systems may again provide a biased vision of the early 
universe, as neither giant starbursts that would be dust obscured, nor
the metal rich
passively evolving spheroids or the hot ICM/IGM enlist among DLAs.
Nevertheless, the average metallicity of the absorbers at $z=3$ appears to be
$\sim 1/20$ solar (Pettini et al. 1997, cf. Fig. 4), just a factor of
2  below the expected value from the {\it fossil evidence}. 
However, this is much higher than the lower limit to $Z$ at $z=3$ as
inferred from Ly$_\alpha$ forest observations (Songaila 1997), which
suggests the universe being very inhomogeneous at that epoch.

I would like to make a last point before closing. While there is now
compelling evidence for the stellar populations of ellipticals and
bulges being very old, it is fair to say that the evidence for rather old
disk is also growing.  As well known, the ratio of present to average
past SFR increases along the Hubble sequence, and is well below  unity
at least up to Sbc galaxies (Kennicutt et al. 1994). This means that
SFRs in such disks were much higher in the past, hence disks are also
dominated by rather old stars. Moreover, all SFR indicators are found to 
anticorrelate tightly with the $H$-band luminosity (hence mass) of spiral
galaxies (Gavazzi et al. 1996), i.e., the heavier a 
galaxy, the lower its current SFR, and the older its stellar
populations. Big disks appear to have burned most of their gas at
early times, while most of the cosmic star formation is now confined
to lesser galaxies. Hierarchical models of galaxy formation do not
naturally account for the angular momentum of real spirals (Navarro \&
Steinmetz 1997). Perhaps adding the additional problem of accounting
for the observed mass-age correlation may help finding the solution.

\noindent
I would like to thank Mauro Giavalisco and Piero Madau for the
entertaining and instructive discussions we had on these matters.

\end{document}